\documentclass[pdflatex,sn-mathphys-num]{sn-jnl}

\usepackage{graphicx}%
\usepackage{multirow}%
\usepackage{amsmath,amssymb,amsfonts}%
\usepackage{amsthm}%
\usepackage{mathrsfs}%
\usepackage[title]{appendix}%
\usepackage{xcolor}%
\usepackage{textcomp}%
\usepackage{manyfoot}%
\usepackage{booktabs}%
\usepackage{algorithm}%
\usepackage{algorithmicx}%
\usepackage{algpseudocode}%
\usepackage{listings}%

\usepackage{amssymb}

\usepackage{graphicx}

\usepackage{color}
\usepackage{array}
\usepackage{hyperref}
\usepackage{ragged2e}
\usepackage{enumitem}

\usepackage[utf8]{inputenc}
\usepackage{csquotes}
\usepackage[super]{nth}
 \usepackage[switch]{lineno} 

\usepackage[switch]{lineno} 
\usepackage{makecell}
\usepackage[colorinlistoftodos,prependcaption]{todonotes}
\hypersetup{
    colorlinks,
    citecolor=blue,
    filecolor=blue,
    linkcolor=black,
    urlcolor=black
}

\newcolumntype{L}[1]{>{\raggedright\let\newline\\\arraybackslash\hspace{0pt}}m{#1}}
\newcolumntype{C}[1]{>{\centering\let\newline\\\arraybackslash\hspace{0pt}}m{#1}}

\newcounter{EKXCommentsCounter}

\newcounter{JGHCommentsCounter}

\begin{document}

\title[Article Title]{When Continuous Delivery Is Not an Option: Practical Paths to Continuous Engineering in Complex Organizations}



\author*[1]{\fnm{Eriks} \sur{Klotins}}\email{eriks.klotins@bth.se}

\author[2]{\fnm{Magnus} \sur{Ahlgren}}

\author[2]{\fnm{Nicolas} \sur{Martin Vivaldi}}

\author[2]{\fnm{Even-André} \sur{Karlsson}}

\affil*[1]{\orgdiv{Software Engineering Research Lab~(SERL)}, \orgname{Blekinge Institute of Technology}, \orgaddress{\country{Sweden}}}

\affil[2]{\orgname{Addalot AB}, \orgaddress{\country{Sweden}}}


\abstract{
  
\textbf{Purpose:} Continuous Software Engineering (CSE) promises improved efficiency, quality, and responsiveness in software-intensive organizations. However, fully adopting CSE is often constrained by complex products, legacy systems, organizational inertia, and regulatory requirements. In this paper, we examine four industrial cases from the automation, automotive, retail, and chemical sectors to explore how such constraints shape CSE adoption in practice. 

\textbf{Methods:} We apply and extend a previously proposed CSE Industry Readiness Model to assess the current and potential levels of adoption in each case. Through expert interviews and narrative synthesis, we identify common driving forces and adoption barriers, including organizational preparedness, cross-organizational dependencies, and limited customer demand for continuous delivery. 

\textbf{Results:} Based on our findings, we propose an updated readiness model that introduces additional levels of internal and external feedback, distinguishes market- and organization-facing constraints, and better guides practitioners in setting realistic CSE adoption goals. 

\textbf{Conclusions:} Our results highlight that while full end-to-end CSE adoption may not always be feasible, meaningful internal improvements are still possible and beneficial. This study provides empirically grounded guidance for organizations navigating partial or constrained CSE transformations.

}

\keywords{Continuous Software Engineering, CI/CD, industry readiness, experience report}



\maketitle

\section{Introduction}


Continuous software engineering (CSE) is gaining traction in the industry. The promised benefits of faster software delivery, improved quality, increased efficiency, and customer satisfaction, among other benefits, are appealing to many~\cite{humble2018accelerate}. However, introducing automation, organizational and technical issues, and misaligned expectations are sizeable challenges hindering organizations from reaching the full potential of continuous engineering~\cite{chen2015continuous,fitzgerald2014continuous,klotins2022towards}.

Companies must fulfill certain prerequisites to have the potential to realize the full benefits of continuous engineering. For instance, they may need revise their ways of working, and the business models to support continuous delivery.~\cite{klotins2023continuous,klotins2022towards}. However, industrial software engineering seldom exists in a bubble. Organizations are entangled in an ecosystem of customers, suppliers, regulators, partners, legacy software, organizational structures, and business models~\cite{manikas2013software}. This ecosystem imposes practical limits on what can and cannot be changed. Thus, there are constraints on to what extent continuous engineering can be adopted and its benefits realized~\cite{klotins2022towards}.

These constraints can hinder the adoption of CSE in two ways. Firstly, organizations may inadvertently attempt to attain benefits that are not feasible, given their specific constraints. For example, investing in attaining the continuous deployment capability for software operated by customers and outside the supplier's control will not bring the expected benefits. This is a typical situation in business-to-business software~\cite{popp2011software}. Secondly, facing some obstacles, organizations may abandon plans to consider continuous engineering altogether, thus missing at least partial benefits. For instance, automated and frequent deliveries may not be feasible, given the lack of control over operations. However, automation and continuous integration practices may yield substantial benefits, nevertheless. 

Importantly, individual stakeholders may not be aware of the complete picture, as constraints may emerge at different organizational levels and outside their purview. For instance, engineers pushing for automation and a new delivery model may not be aware of customers' attitudes toward changing the model. Similarly, upper management strategizing to improve software delivery performance, may not be aware of the underlying technical challenges.

State-of-the-art emphasizes the benefits of automation in software engineering, such as, testing, builds, configuration management, integration, and other routine tasks~\cite{rafi2012benefits}. An end-to-end automated software delivery pipeline enables frequent and automated software deliveries to customers (CICD), extending the benefits with frequent value delivery~\cite{fitzgerald2017continuous,humble2018accelerate}.

Literature so far mostly considers technical limitations (e.g., inadequate architecture~\cite{bril2000maintaining,chen2018microservices}, adoption of automation~\cite{nass2021many}) and organizational constraints (e.g., culture and attitudes~\cite{chen2015continuous,klotins2023organizational}) that must be altered to attain the benefits. Such a perspective misses the broader view of constraints that shape organizations and affect their flexibility to adopt CSE practices. 
We also observe that literature mostly contains success stories of CSE adoption. Such a one-sided view fails to explore and analyze lessons learned from complex cases where CSE adoption is problematic.

Our earlier work proposed the industry readiness model for adopting CSE
~\cite{klotins2023continuous}. 
The model attempts to structure the prerequisites, challenges, and likely benefits of adopting CSE practices. The pyramid shape illustrates how the foundation of organizational readiness enables the adoption of automation on various levels. Once automation and customer agreements are in place, the automation can be extended to enable continuous solution delivery, paving a path towards continuous collection and use of data and evolving value streams.

In this study we explore four industrial cases and their CSE adoption efforts. The studied cases are large companies with complex software and present interesting cases for studying CSE adoption challenges. We use the industry readiness model to analyze the prerequisites and discuss specific CSE adoption constraints. In the end, we discuss the model's usefulness in evaluating industrial readiness for CSE.

This study makes several contributions to state-of-the-art. Firstly, we summarize the state of practice for CSE adoption in four industrial cases. The cases represent large organizations, complex products, and challenging adoption of CSE practices. Hence, we improve and address the gap in state-of-the-art by adding lessons learned from these complex cases. Secondly, we demonstrate the application of the CSE industry readiness model to determine the optimal level of CSE adoption in the industry. Setting realistic adoption goals is important to maximize benefits from at least partial adoption of CSE and to avoid wasted resources on pursuing unattainable benefits. Lastly, we use the lessons from the studied cases to validate and improve the industry readiness model.

The rest of this paper is structured as follows. Sections 2 presents background of lean, agile and continuous software engineering, as well state-of-the-art in the adoption of agile methods in safety critical domains. Section 3 outlines our research methodology. Section 4 presents the studied cases and analysis of their CSE adoption efforts. Section 5 discusses our findings, and Section 6 concludes the paper.

\section{Background and related work}

\subsection{Lean, agile, and continuous software engineering}

Continuous software engineering (CSE) originates from lean and agile principles in software engineering. The principles emphasize value delivery, anticipation of change, flexibility, and waste minimization. CSE implements these principles by delivering software incrementally and frequently. With the help of automated tools, test suites, and staging environments, CSE enables organizations to ship the most recent software changes several times daily~\cite{fowler2001agile,rodriguez2012analyzing,chen2015continuous}. 

\begin{figure*}[ht!]
  \centering
  \includegraphics[width=\textwidth]{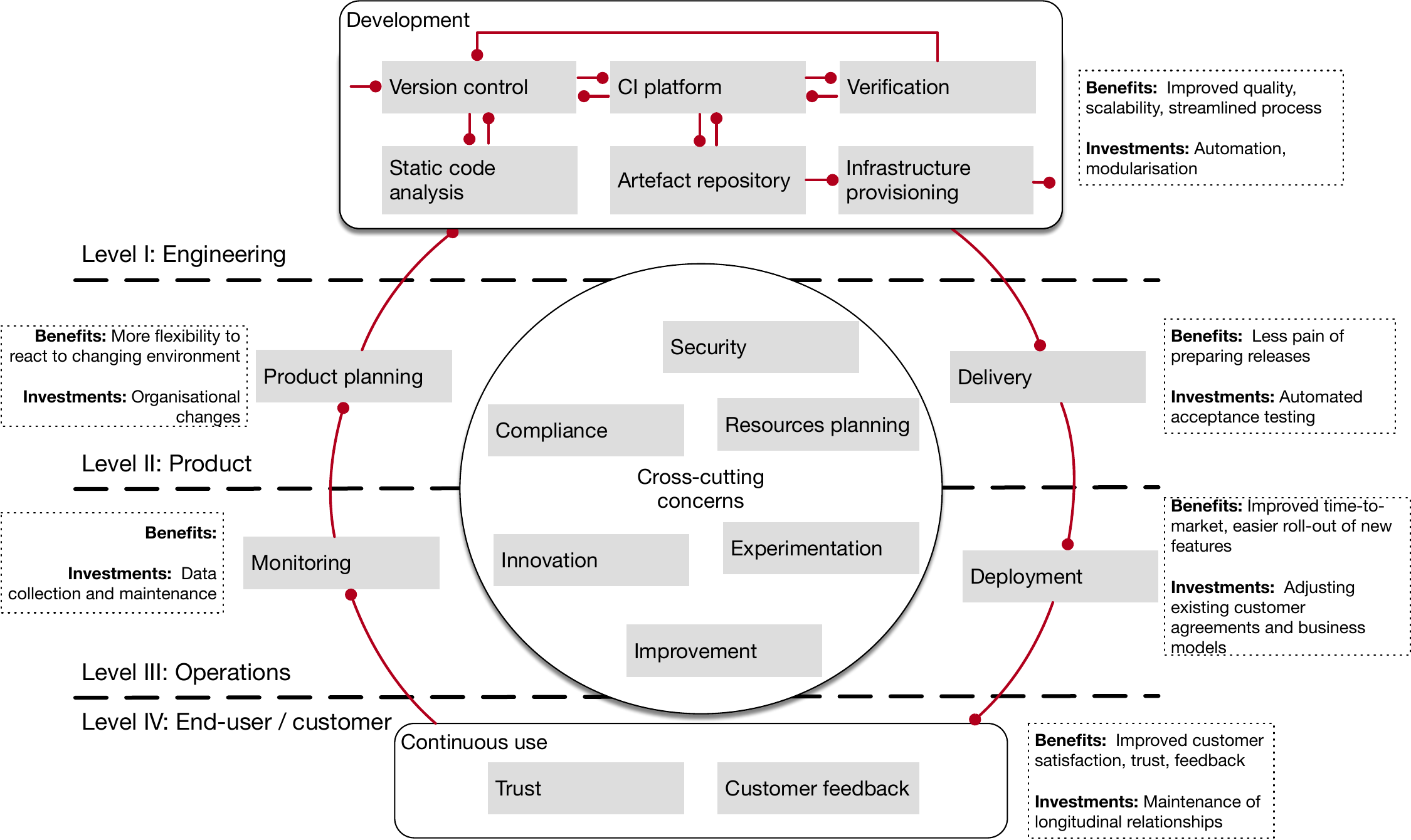}

  \caption{The conceptual mode of continuous software engineering (adopted from Klotins et al.~\cite{klotins2022towards})} 
  \label{fig:conceptual_model}     
\end{figure*}

The CSE primarily comprises of connected and streamlined automated build, integration, verification, and delivery steps, i.e. CI/CD~\cite{poth2018deliver}. However, we consider a broader perspective of CSE that includes a closed feedback loop, see Fig.~\ref{fig:conceptual_model}.

The benefits of CSE are diverse. Firstly, frequent and small deliveries enable precise and almost immediate customer feedback. Frequent feedback enables continuous planning, allowing organizations to minimize the gap between what is delivered and what customers expect. Secondly, frequent deliveries are aimed to mitigate the risk and disruption caused by large software updates. Smaller changes are relatively easier to develop, integrate, verify, and deploy, and the disruption is limited in case of a failure. Thirdly, frequent deliveries to customers and their feedback positively affect engineers' motivation, job satisfaction, and reduce stress~\cite{humble2018accelerate,chen2015continuous}. 

Implementing CSE in an organization is a sizeable undertaking. The organization must adopt specific software architectures, tools, and practices, as well as develop and maintain the toolchain, test suites, and test data, among other technology concerns. Organizational structures, culture, and ways of working must embrace CSE principles~\cite{fitzgerald2017continuous,klotins2022towards,chen2017continuous}.

Retrofitting an existing product and organization with CSE is an even more substantial challenge. New tools and ways of working must not only be implemented but must also be implemented without disrupting ongoing operations and customer contracts. In our earlier work, we explore how existing organizations may not be able to follow through with the necessary changes due to business risks, internal and external constraints, and unclear investment/benefit perspectives
~\cite{klotins2022towards,chen2017continuous,klotins2023organizational}.

\subsection{The adoption of agile methods in safety-critical domains and complex organizations}
The literature on implementing CSE in complex and regulated domains is scarce. Yet, CSE shares many characteristics with agile development methodologies, such as emphasis on value delivery, iterative work, and lightweight documentation.

Heeager et al.~\cite{heeager2018conceptual} propose a model of agile development in a safety-critical context and identify four key problem areas:

\begin{enumerate}
    \item Requirements flexibility hindering traceability and safety

    \item  Iterative and incremental lifecycle complicating functional, non-functional, and compliance verification and product validation

    \item Lightweight documentation hindering traceability and verification

    \item Test first approach incompatible with regulatory compliance

\end{enumerate}

The identified issues point towards the conflict between flexible and lightweight agile development practices and rigorous compliance requirements. Such challenges are very relevant for adopting CSE.

Kalenda et al.~\cite{kalenda2018scaling} review the challenges and success factors of adopting agile methodologies in large organizations. The authors list resistance to change and lack of knowledge as the most frequently reported challenges. At the same time, acquiring knowledge, careful transformation, and executive sponsorship are listed as success factors.

State-of-the-art in adopting agile methods in large and complex organizations resonates well with the literature on CSE transformations~\cite{luz2019adopting,klotins2022towards,chen2015continuous}. Support throughout the organization, knowledge of modern software engineering methods, and well-planned transformation are crucial. Moreover, agile and CSE methods need to be tailored to fit specific organizational context~\cite{rolland2016tailoring}.

\subsection{Industry readiness for continuous software engineering}

Analysis of how established organizations can adopt CSE led us to formulate the industry readiness model. The model formulates levels of CSE adoption along with preconditions, practices, showstoppers, and a discussion of what level is appropriate for an organization~\cite{klotins2023continuous}.

\begin{figure*}[ht!]
  \centering
  \includegraphics[width=\textwidth]{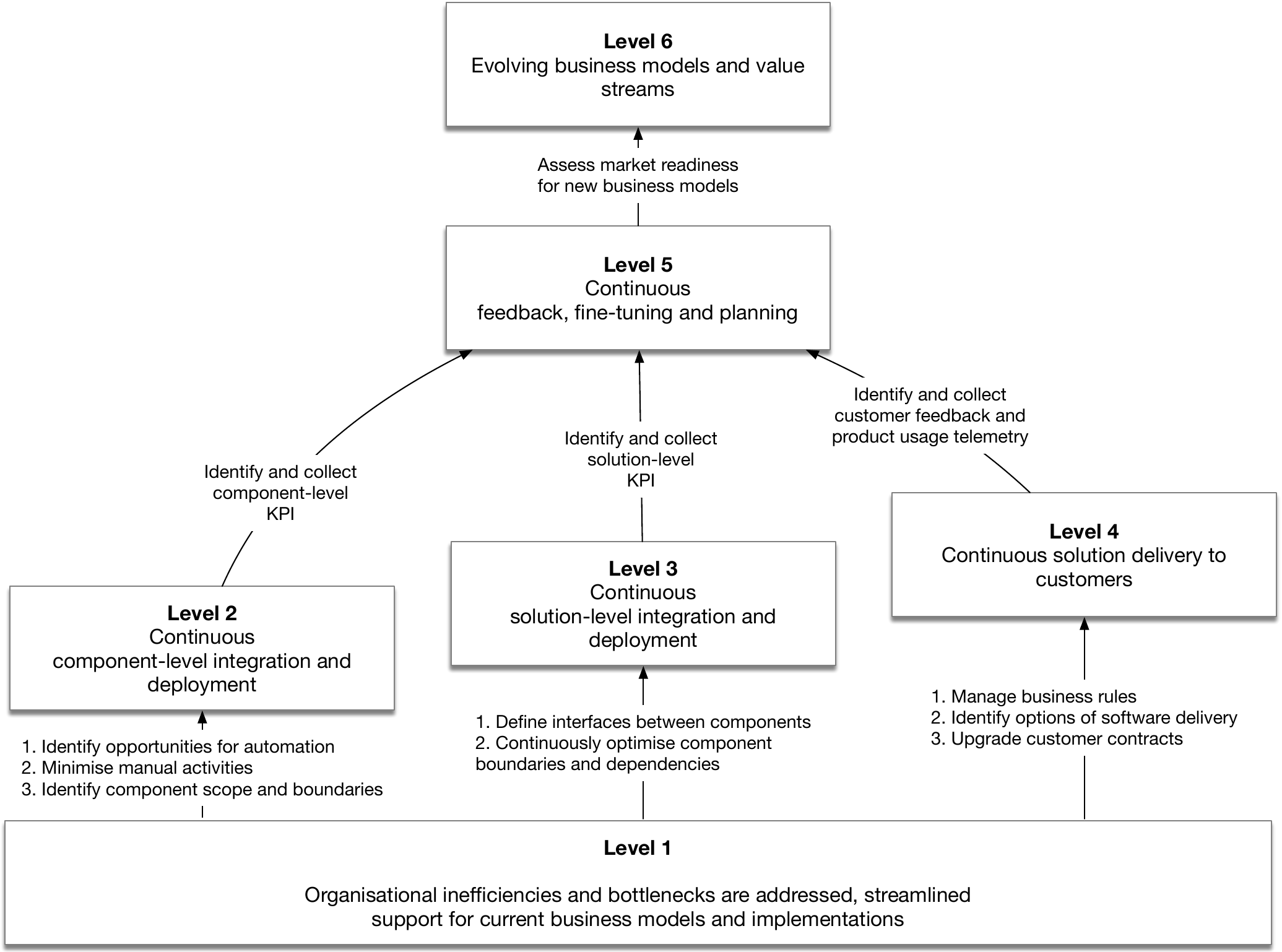}

  \caption{Industry readiness model (adopted from Klotins et al.~\cite{klotins2023continuous})}
  \label{fig:rediness_model}     
\end{figure*}

The readiness model, see Fig.~\ref{fig:rediness_model} identifies 6 levels of CSE adoption. Moving up the levels requires meeting certain prerequisites, and the levels build upon each other.

\begin{enumerate}
  \item \textbf{Organizational readiness} expressed as relative maturity in current ways of working, stakeholder support and clear adoption goals, is required to undertake further steps.

  \item \textbf{Component-level integration and deployment} permitting isolated islands of automation can be achieved once the organization defines the scope and low-hanging fruits for automation.

  \item \textbf{Solution level integration and deployment} permitting automated builds of the whole solution can be achieved once process interfaces are defined and individual islands of automation connected.

  \item \textbf{Solution delivery} enabling automated and frequent deliveries of software to end users can be attained once permitted by the business model and customer service level agreements.

  \item \textbf{Continuous feedback and fine-tuning} can be achieved once user and process telemetry is collected and analyzed to fine tune both the product and its delivery.

  \item \textbf{Evolving business models and value streams} arise from the earlier levels once the organization can move away from transactional software delivery model to a more collaborative service model.
\end{enumerate}


We observe a survivor's bias in publications about CSE. Cases where CSE adoption is successful gain more attention and reporting in literature. However, cases where the adoption of CSE is unsuccessful, stalling, or canceled altogether attain limited attention. However, the latter could present valuable lessons learned for both CSE state-of-the-art and practice.

\section{Research Methodology}

\subsection{Research aims}
This study aims to extend and validate our earlier work, gouging the right level of CSE adoption We focus on particularly challenging cases where end-to-end adoption of CSE is not feasible. Yet, a limited adoption of CSE could yield significant benefits.

We guide our study with the following research questions:
\vspace{0.5cm}

\noindent\textbf{RQ1:} What is the state of practice in adopting CSE?

\noindent\textit{Motivation:} With this research question, we aim to explore the current status, challenges, constraints, and ambitions towards CSE in each studied case.
\vspace{0.3cm}

\noindent\textbf{RQ2:} What driving forces and constraints determine the optimal CSE adoption level?

\noindent\textit{Motivation:} With this research question, we aim to identify factors that contribute to or hinder the adoption of CSE. 
\vspace{0.3cm}

\noindent\textbf{RQ3:} To what extent is the CSE industry readiness model useful for setting goals and identifying constraints in the adoption of CSE?

\noindent\textit{Motivation:} With this research question, we aim to map the state of practice (RQ1) and relevant factors (RQ2) to the CSE adoption model and validate the usefulness of the model to guide further CSE adoption.

\section{Research approach}

We use a qualitative research approach and narrative synthesis to answer our research questions and attain the aim of our study~\cite{huang2018synthesizing,popay2006guidance}. Qualitative research holistically explores a phenomenon in its context. We opt for narrative synthesis to capture the rich experience of the interviewed experts and capture nuances of each case. We further apply our industry readiness model~\cite{klotins2023continuous} as a lens to further analyze each case and to highlight commonalities and differences in relation to CSE adoption barriers.

In our study, the CSE is a relatively well-defined concept. Earlier work focuses mainly on the elements and mechanics of CSE~\cite{fitzgerald2017continuous,kim2008automated,humble2018accelerate}. However, the context of CSE adoption and any contextual success factors and hindrances are less explored. Hence, a holistic qualitative approach is required to explore the context of CSE adoption.

Our study primarily uses semi-structured interviews with industry practitioners and two-way discussions to identify company ambitions and context, product specifics, current ways of working, and CSE adoption. The data collection was conducted in three steps:

\subsection{Step 1 - Planning and introduction}

The first step of our study was to identify suitable experts and cases. We contacted our industry collaboration partners and identified experts with experience in leading CSE adoption efforts. We further selected particularly interesting cases from a contextual challenges perspective. That is, we searched for complex, regulated products in large organizations seeking to benefit from continuous practices, however cannot adopt CSE by the book. 

We summarize the list of experts and corresponding cases in Table~\ref{table:experts_summary}. To protect the privacy of our experts and the companies, we have anonymized the cases and provided only the product description and context.

\begin{table*}[!th]
  \renewcommand{\arraystretch}{1.2}
  \caption{The summary of study participants and the studied cases}
  \label{table:experts_summary}
  \centering
  \begin{tabular}{L{0.6in}L{1.5in}L{1.7in}L{0.6in}}
  \hline
     Expert & Role & Case description & Time with the case \\

    \hline
      Expert 1 & Process quality expert & Industrial automation supplier & 4 years \\
      Expert 2 & Functional safety and cybersecurity expert & Tier 1 automotive supplier & 6 years \\
      Expert 3 & Project manager, agile leader & E-commerce department in a large brick and mortar store chain & 1 year \\
      Expert 4 & Project manager, agile leader & Supply chain management software for a global chemical company & 5 years \\

    \hline
 \end{tabular}
 \end{table*}

As part of the introduction, we organized a two-part workshop. The first part focused on establishing a common terminology and a frame of reference for further discussion. We presented and discussed CSE in general, adoption challenges, and lessons learned from state-of-the-art. 

In the second part, the experts presented their cases, associated challenges, lessons learned, and interesting analysis points to explore further. 

The workshops took 1 hour and 3 hours for each part, respectively. The presentations, whiteboard drawings, and workshop notes were collected and archived for future reference.

\subsection{Step 2 - Expert interviews}
We followed by conducting interviews with each expert separately to gain more in-depth insights about each specific case. Based on insights from Step 1, we formulated an interview guide, see Table~\ref{table:expert_interview_guide}. Each interview took about an hour and was conducted online.

\begin{table*}[!th]

  \renewcommand{\cellalign}{lc}
\renewcommand\cellgape{\Gape[3pt]}
  \caption{Expert interview guide}
  \label{table:expert_interview_guide}
  \centering
  \begin{tabular}{L{0.2in}L{1in}L{3.5in}}
  \hline
     \# & Focus area & Questions (exemplified)\\

    \hline
    1 & The expert & \makecell{How long have you been with the company/case?\\What is your role/position/responsibilities?\\How would you describe your experience with CSE?} \\
    2 & The company & \makecell{What is the size of the company (product team)?\\What is the business model?\\What is the product domain?\\What are the markets in which the company operates?}\\
    3 & The product & \makecell{What is the product?\\How would you characterize the size of the product?\\Is the product regulated, business, or safety-critical?\\What is the degree of control over the upstream\\ and downstream supply chain?}\\
    4 & Current process & \makecell{Walk me through the steps of the software delivery\\process, from requirements to product delivery.\\What are the dependencies on third parties?}\\
    5 & Driving forces & \makecell{Why are you considering CSE?\\What benefits do you aim to attain?\\What goals drive the adoption of CSE?\\From what level in the organization does the CSE\\initiative originate? How broad is the support?}\\
    6 & Challenges & \makecell{What challenges (organizational, technology, regulatory,\\customer-related) have you experienced?\\What prevents you from adopting CSE by the book?}\\
    7 & Best realistic outcomes & \makecell{In your view, what is the best realistic\\outcome from CSE adoption in your organization? Why?}\\
    8 & Lessons learned & \makecell{What are the key lessons learned from working\\ on CSE adoption}\\

    \hline
 \end{tabular}
 \end{table*}

\subsection{Part 3 - Analysis}

The analysis of interviews started already during the interview process where we asked the experts to reflect, summarize, and connect their insights to state-of-the-art. Each interview provided ideas and new analysis points for the upcoming interviews. This way, we gathered broad expert views on the adoption of CSE and were able to clarify and connect lessons learned.

After the interviews, the lessons learned, challenges, and analysis points were mapped to the CSE industry readiness model. With the mapping, we aim to identify overlapping concepts, that is, challenges and adoption constraints already in the model. Thus, adding more details and validation to the model. We further look for additional constraints and challenges that are currently not part of the CSE readiness model. The new insights were considered for further improving and developing the readiness model. After the analysis, the experts reviewed the results for validation and further interpretation. 

For each case we evaluate the predicted (by our earlier readiness model) and observed (by the expert judgement) CSE readiness level. We differentiate between current, potential, planned, failed, not feasible, and irrelevant adoption stages. We summarize them in Table~\ref{table:adoption stages}.

\begin{table*}[!th]
  \renewcommand{\arraystretch}{1.2}
  \caption{CSE level adoption stages}
  \label{table:adoption stages}
  \centering
\begin{tabular}{L{1in}L{3.7in}}
  \hline
     Stage & Interpretation\\
    \hline
      
      Current & The case has already attained the specified level at the time of this study. \\

      Potential & The case demonstrates the capacity to reach the specified level under favorable conditions. \\

      Planned & The team or organization behind the case has concrete plans to attain the specified level in the near future. \\

      Failed & Attempts to attain the specified level were unsuccessful and have been discontinued. \\

      Not feasible & Due to internal constraints, attaining the specified level is not realistically achievable. \\

      Irrelevant & The specified level is not considered meaningful or valuable for the case, given external conditions.\\

    \hline
 \end{tabular}
 \end{table*}

\subsection{Threats to validity}
In this section we present and discuss potential shortcomings and limitations of our study. We follow the structure suggested by Wohlin et al.~\cite{wohlin2012experimentation}:

\subsubsection{Internal validity}

Internal validity concerns the causal relationships between observations and outcomes. A potential threat in this study arises from the dual role of the experts as both informants and co-authors. While such setup provided deep insights and accurate contextualization, it may have introduced confirmation bias, as experts might have unconsciously downplayed organizational weaknesses or overemphasized progress, and improperly identified the causes and effects. We mitigated this risk through iterative triangulation, in which early interview insights were revisited in subsequent sessions, contrasting viewpoints were explicitly discussed.

 \subsubsection{Construct validity}

Construct validity addresses how well the concepts under study are captured by the research design. The central construct in this study, the level of CSE adoption-—was assessed using the CSE Industry Readiness Model. While this model provides a structured lens for evaluation, its application depends on subjective interpretation of expert statements and organizational artifacts.

To mitigate this threat we started the expert interviews with a presentation of the key terms and concepts, followed by a semi-structured interview allowing the interviewees to describe the case without any specific lens. Then during the analysis process, we mapped their statements to the industry readiness model, and conducted additional review rounds to validate our assessment.

\subsubsection{External Validity}

External validity refers to the extent to which findings are generalizable to other contexts. The selected cases span multiple domains (industrial automation, automotive, retail, chemicals), but they all come from large, well-established companies operating in regulated or complex environments. As such, the insights may not directly apply to start-ups, purely digital-native firms, or unregulated domains, where CSE adoption dynamics may differ significantly.

\subsubsection{Reliability}
Reliability concerns the consistency and repeatability of the study's procedures and findings, and independence from researcher biases. The use of semi-structured interviews, a standardized expert interview guide, and structured readiness level assessments ensures research process transparency.

Further, the involve experts were invited to contribute and revise the manuscript thus triangulating and providing a degree of quality control over the interpretations and conclusions.

\section{Results}

\subsection{Case I -- Industrial Automation Supplier}

Case I represents an industrial automation supplier. The automation solutions contain both software and hardware to be used in factories, power plants, and alike. The solutions are used in safety and business-critical environments, thus following strict regulations and certifications. 

The company (Case I) develops standardized components and provides services to install and maintain the systems. The components are installed on sites in bespoke configurations to fit the facility's requirements. The installation comprises of hardware, software, and the configuration of software.

The solution forgoes two stages of quality assurance. In the first stage, the company uses several reference solutions, i.e., mock facilities, to perform solution-level verification in-house. However, some functionalities and quality aspects can only be verified after the final installation on site. Post-deployment validation and certification provide assurance of the system’s safe operation within the facility. On-site verification is an effort-intensive process involving software, hardware, machinery in the facility, and third-party quality auditors.

Customers have a limited interest in new features once the automation system is installed and certified. Firstly, the regulated and safety-critical nature of the solution imposes constraints on what can be implemented. Secondly, the software works in tandem with the hardware in the facility. Thus, any software updates likely depend on hardware updates. Thirdly, the long lead times and costs of hardware updates, systems installation, and certification limit the potential of frequent software deliveries. Thus, frequent deployment of new features is not considered a viable goal.

The aims towards the adoption of continuous engineering are limited to streamlining and automating build, integration, and in-house verification processes. Currently, the company is exploring continuous integration and low-level test automation to facilitate learning, improve internal efficiency, and prevent bug slippage.

There is clear management support for CSE in Case I, and the company is primarily interested in the internal benefits, such as, improved efficiency, traceability, transparency, and quality. 

The adoption of CSE is hindered by low market pull for new features and low customer interest in upgrades. Difficulties in continuous delivery limit how much telemetry and feedback can be collected. Thus, introducing new organizational structures and ways of working requires a significant effort.

\subsubsection{Readiness assessment}

In our readiness assessment we identify the current, target, and potential CSE adoption level. We use the levels and follow the guidelines from our earlier work
~\cite{klotins2023continuous} 
and summarize our assessment in Table~\ref{table:lvls_case_1}. In the table, we analyze the adoption levels, predicted adoption level according to the case details, observed adoption level based on experts' insights, and notes concerning the assessment.

\begin{table*}[!th]
  \renewcommand{\arraystretch}{1.2}
  \caption{CSE readiness level analysis of Case I}
  \label{table:lvls_case_1}
  \centering
\begin{tabular}{L{1.3in}L{0.81in}L{0.81in}L{1.7in}}
  \hline
     Level & Predicted & Observed  & Notes\\

    \hline
      1~Org. readiness              & Potential & Current & The organization has clear goals and ambitions\\ 
      2~Component-level automation  & Potential & Current & Automation initiatives are underway\\
      3~Solution-level automation   & Potential & Planned*& Possible only to reference solutions \\
      4~Continuous delivery         & Not feasible & Not feasible & Not feasible due to safety regulations\\
      5~Continuous data             & Not feasible & Potential*& Limited to continuous process data\\
      6~Evolving business           & Not feasible & Irrelevant & Not relevant due to domain/market specifics\\
     
    \hline
 \end{tabular}
 \end{table*}


     

We estimate that Case I has reached the organizational readiness level (Level 1). The product in question, the customer agreements, and the business model are relatively stable and mature. The organization has defined its ambitions and limitations to CSE adoption. 

Levels 2 and 3 are within reach for Case I, see Table~\ref{table:lvls_case_1}. Currently, the organization works towards component-level automation. The automated pipeline can be extended to the whole solution level, albeit limited to reference solutions. Reaching Level 4, continuous software delivery to facilities, is not feasible due to safety regulations and the critical nature of the product. 

Continuous use of data, Level 5, is partially attainable. Although live product telemetry is out of the question, process data can still be used to fine-tune the engineering process and inform Levels 2-3.

Level 6 of evolving business models is currently not prioritized in Case I as the company, the product, and the entire domain are regulated, relatively well established, and moving slowly.

\subsection{Case II -- Automotive parts provider}

Case II is a Tier 1 automotive parts provider. In our study, we are considering the unit working on developing a drive train component. The component comprises two key parts. The mechanics and a
control unit (ECU) allowing dynamic adjustments on the performance. The software in the ECU is the focus of our investigation. The size of the software is in the ballpark of 100k LOC. The vendor (Case II) is a supplier to other automotive manufacturers who install the component in their vehicles.  

Due to automotive regulations (UN ECE), the provider must be able to update the unit up to 10 years after the last car manufacturing date. For instance, when critical defects are discovered.
This leads to extremely long maintenance cycles and the need to support outdated product versions. For instance, a control unit may be in development for 5 years, a car may be in production for 10 years, and the provider must be able to reproduce the control unit for 15 more years. This leads to the need to maintain the capacity to support and produce 25+ years old software and hardware.

The need to support long maintenance cycles leads to difficulties with tooling and automation. For instance, relevant compilers, libraries, development environments, and operating systems may be obsolete and no longer maintained. The hardware to run such tooling may be no longer available. Modern virtualization, containerization, and automation tools poorly support such outdated software. Thus, the company gets creative in ensuring compliance. For instance, keeping an old Windows XP machine on a shelf if they are ever required to rebuild the old software. Furthermore, engineers familiar with such technologies, tools, and know-how may no longer be around, further exacerbating the challenges with long-term software maintenance.

Car manufacturers have slightly different requirements for the driving experience and the component's performance. This is achieved by offering slightly customized versions to different manufacturers. The final modifications of the component design are made during a collaborative process and before a car model goes into production. Once a car goes into production and units are installed on vehicles, there are limited means for the vendor to connect, update, or extract data from the component. Extracting data requires either not yet broadly adopted over-the-air connectivity or data extraction must be done at a specialized workshop. 

Furthermore, even the means to update car components over the air may exist, the car maker needs to want and enable it. Some car manufacturers, Tesla being the most prominent example, are updating their car software often. However, others are still lagging behind and does not practice frequent software updated for different reasons.  

The current ambitions towards continuous engineering are limited to automating and streamlining the in-house development, build, and verification process. As well as improving the speed of delivering changes between the case company and car manufacturer during the component customization process. However, due to the downstream dependencies on other car manufacturers, pushing for continuous deliveries is not considered.
The new feature requests typically concern mechanical features or adjusting the working range of the component. Such requests are typically well-known long in advance. Unique and innovative features are rare.

The expert pointed out an increasing trend from car manufacturers to take over the control over driving experience, and by extension have a more granular control over the component behavior. The vendor (Case II) used to provide some logic in the ECU to control the driving experience; however, the recent models provide only a basic integration interface, leaving the control over the driving experience to the car manufacturer's software. The rationale for this trend is the increasing physical and logistical complexity of providing a consistent experience from different components provided by different suppliers. To streamline the experience, car manufacturers move such functionality centrally. Furthermore, the wish to control more aspects of the software supply chain is inspired by Tesla. The complete control over the car's software and hardware enables Tesla to roll out changes frequently and offer an ever improving driving experiences. Such trends offer an opportunity to work more closely with the car manufacturers and rethink the distribution of responsibilities between parts suppliers and car manufacturers.

\subsubsection{Readiness assessment}

In our readiness assessment we identify the current, target, and potential CSE adoption level. We use the levels and follow the guidelines from our earlier work
~\cite{klotins2023continuous} 
and summarize our assessment in Table~\ref{table:lvls_case_2}.

\begin{table*}[!th]
  \renewcommand{\arraystretch}{1.2}
  \caption{CSE readiness level analysis of Case II}
  \label{table:lvls_case_2}
  \centering
  \begin{tabular}{L{1.3in}L{0.81in}L{0.81in}L{1.7in}}
  \hline
     Level & Predicted & Observed  & Notes\\

    \hline
      1~Org. readiness              & Potential & Current & The organization has clear goals and ambitions\\ 
      2~Component-level automation  & Potential  & Ongoing & Automation initiatives are underway\\
      3~Solution-level automation   & Potential  & Planned*& Development tool chain automation \\
      4~Continuous delivery         & Not feasible & Not feasible & Impossible due to the air gap\\
      5~Continuous data             & Not feasible & Potential*& Limited to continuous process data\\
      6~Evolving business           & Not feasible & Potential* & Markets demands are changing and slowly creating new opportunities\\
     
    \hline
 \end{tabular}
 \end{table*}

Similarly to Case I, the organization and the product in Case II are well established with clear ambitions towards improving software delivery. Hence, the Case II has achieved Level 1 adoption. Component and Solution level automation (Levels 2-3) are currently in progress; see Table~\ref{table:lvls_case_2}.

Continuous software deliveries to end users (Level 4), i.e. cars on the road, are not considered for several reasons. Firstly, the infrastructure to access the vehicle and update its software over the air is limited to most recent models. Secondly, the car manufacturers take downstream responsibility for the installation and maintenance of the component. Thus, the company in Case II has no practical control over the software once a car leaves the factory.  Due to limited access, continuous collection of product usage data (Level 5) is out of the question once installed. However, continuous use of process data is very much attainable. 

Evolving business models (Level 6) are currently not considered. However, as pointed out by the expert, car manufacturers are moving towards the Tesla model of closer integration of different ECUs and improved driving experience. The improvements in driving experience are largely based on sharing driving data with the manufacturer and frequent software updates to the vehicle. Such a paradigm shift in the automotive domain may open up opportunities for new business models and value streams. 

\subsection{Case III -- Global Retailer}

Case III is an IT unit developing and maintaining an online e-commerce platform for a global retailer of consumer products and operates in a large number of countries. Historically, the company prioritized brick-and-mortar stores to provide a unique shopping experience. However, recently the e-commerce operations have grown substantially. They have a large number of IT/R\&D employees managing their IT systems. 

The company has an extensive network of partners and suppliers to operate global brick-and-mortar stores. Each partner has its own IT strategy and infrastructure. Orchestrating the interfaces with partners' systems and ensuring accurate supply chain management is a major undertaking.

We focus our inquiry on a team of 15-20 engineers working on a customer-facing e-commerce system. The e-commerce system is developed in-house, tightly coupled, and integrated with 3rd party systems.

The company behind Case III is very cost-conscious, and there is a constant push to improve the efficiency of IT operations. This push has led to a recent transformation from plan-driven development to adopting agile and, most recently, CSE methodologies. 

Customer satisfaction, consistent shopping experience online and in brick-and-mortar stores, and rapid evolution of e-commerce platforms are the key drivers for considering adopting continuous engineering in the organization. The high expectations of the e-commerce platform and a relatively small team enabled a pocket of excellence in terms of adopting continuous engineering. The rest of the organization lags as the pockets of excellence are difficult to scale. The challenges to scale the good initiatives are multi-faceted and concern the scale of the organization, third party-suppliers, culture, technology, and internal resistance.

Installing continuous engineering in a large and mature organization takes time. Supply chain systems connecting  a great number of physical and online stores cannot be changed easily. The systems are interdependent, business-critical, and do not share the same pace of evolution. Moreover, these systems are tied to many external systems, different legislations, regional differences, service level agreements, hardware, and working methods. 

With a long history of providing unique shopping experiences in brick-and-mortar stores, the 
organization's culture changes slowly. E-commerce offers a fundamentally different type of shopping experience. The required culture shift takes time and creates organizational friction to embrace the rapid evolution of e-commerce systems.

That being said, the nature of e-commerce systems provides the necessary enablers for CSE, for instance, the of ownership of the delivery pipeline with better automation capabilities than much of the other parts of the legacy environment.

\subsubsection{Readiness assessment}

In our readiness assessment we identify the current, target, and potential CSE adoption level. We use the levels and follow the guidelines from our earlier work
~\cite{klotins2023continuous} 
and summarize our assessment in Table~\ref{table:lvls_case_3}.

\begin{table*}[!th]
  \renewcommand{\arraystretch}{1.2}
  \caption{CSE readiness level analysis of Case III}
  \label{table:lvls_case_3}
  \centering
\begin{tabular}{L{1.3in}L{0.81in}L{0.81in}L{1.7in}}
  \hline
     Level & Predicted & Observed  & Notes\\

    \hline
      1~Org. readiness            & Potential & Current & The organization has clear goals and ambitions to optimize software delivery\\ 
      2~Component-level automation& Potential & Planned & Automation initiatives are underway\\
      3~Solution-level automation & Potential & Planned & Currently implemented in isolated pockets of excellence \\
      4~Continuous delivery       & Potential & Planned & Possible where the org. has full control \\
      5~Continuous data           & Potential & Potential & Possible, but not considered\\
      6~Evolving business         & Potential & Potential & Shift from physical to online business\\
     
    \hline
 \end{tabular}
 \end{table*}

Case III differs from other cases by their degree of control over the software delivery. The company controls the delivery of the e-commerce solution end-to-end. However, internal dependencies and interfaces hinder the CSE adoption effort.

By our assessment, the Case III has all the prerequisites to reach Level 6 of CSE adoption. E-commerce domain is developing rapidly and company could benefit significantly from the capability to follow the market trends. 

Internally, the company have limited their ambitions to the continuous software deployment. The experts did not mention any initiatives to leverage data (Level 5) in continuous improvement, or an ambition to evolve the e-commerce business models (Level 6).


     

\section{Case IV -- Chemical supplier}

Case IV in our study is a chemical company producing chemicals for end-user applications. The company uses a network of intermediary distributors to deliver their products and services to customers. The end-customer services are primarily supply chain management software and various supporting services, such tools to optimize the procurement and use of their products 

Importantly, the company has a long history and culture of their business is conservative, primarily focusing on chemical products and less on developing customer relationships and exploring new value streams. Historically, the objective of software has been to support the core chemical business. Yet, the importance of software in further enhancing the core business is growing. Furthermore most of the software solutions in Case IV are developed by 3rd parties and operated on-premises. The company depends on many consultants and suppliers to provide software solutions and on in-house engineers to operate the software.

The main driving forces to consider continuous engineering in Case IV are cutting costs, increasing efficiency, and becoming more innovative. Importantly, all know-how and agency over the software solutions lies with the third parties, significantly inflating the costs and hindering internal innovation.

The challenges with software solutions were further exacerbated by the complex network of IT suppliers, chemical distributors, and unclear boundaries of responsibility. Furthermore deploying solutions from IT providers into operation in-house is also a cumbersome process. 

The company initiated an investigation to optimize IT operations by moving IT systems from in-house to cloud environments and delegating more responsibilities to the suppliers. The aim was to reduce operational costs and adopt DevOps by minimizing the split between development and operations environments. Such a move was intended as the first step towards optimizing focus areas and boundaries of responsibility among the chemical supplier and various IT providers.

The investigation did not result in establishing a subsequent implementation project. There were many business reasons for this.  One important reason is organizational preparedness. Optimizing software systems would potentially make several roles and departments obsolete or would lead to change of job description for several employees. The internal resistance hindered a meaningful change process. This strongly contributed to the organization’s decision not to go through with the optimization program.


\subsubsection{Readiness assessment}

Case IV differs from the other cases as it is in the middle between consultants and 3rd party software vendors and the end users. In a way, Case IV is at the receiving end of the software delivery pipeline. 

In Case IV, software delivery is not perceived as part of the value delivery chain. Hence, we evaluate that the company has still to attain organizational readiness (Level 1) and include CSE as part of its software strategy. As demonstrated by the earlier unsuccessful DevOps adoption initiative, the organizational culture and resistance to change prevent meaningful changes in their ways of working. The hesitation to adopt new ways of working and become more open excludes the use of data and evolving business models (Levels 5-6) from consideration; see Table~\ref{table:lvls_case_4}

Nevertheless, there is potential to streamline the relationship between the company in Case IV and various software suppliers. For instance, by establishing a CICD pipeline between software vendors and Case IV (Levels 2-3). However, implementing such a pipeline would require a new service level agreement between vendors and the company regulating continuous software deliveries (Level 4).

\begin{table*}[!th]
  \renewcommand{\arraystretch}{1.2}
  \caption{CSE readiness level analysis of Case IV}
  \label{table:lvls_case_4}
  \centering
\begin{tabular}{L{1.3in}L{0.81in}L{0.81in}L{1.7in}}
  \hline
     Level & Predicted & Observed  & Notes\\

    \hline
      1~Org. readiness            & Potential & Failed &  The org. culture incompatible with DT/CSE\\ 
      2~Component-level automation& Not feasible & - & Outsourced to IT suppliers\\
      3~Solution-level automation & Not feasible & - & Outsourced to IT suppliers \\
      4~Continuous delivery       & Not feasible & - & SLA with external IT suppliers \\
      5~Continuous data           & Not feasible & - & The org. culture incompatible\\
      6~Evolving business         & Not feasible & - & The org. culture incompatible\\
     
    \hline
 \end{tabular}
 \end{table*}


     

\section{Discussion}


\subsection{RQ1: What is the state of practice in adopting CICD?}


In this study we have explored cases where the adoption of CSE was challenging from the beginning. All studied cases are established companies with products already in the market and with relatively successful software delivery practices. Yet, the companies aim to adopt continuous practices to address the growing pressure to optimize costs and to leverage software to increase customer value. We summarize the CSE adoption levels in all studied cases in Table~\ref{table:lvls_case_1-4}.

\begin{table*}[!th]
  \renewcommand{\arraystretch}{1.2}
  \caption{The summary of CSE readiness levels of the studied cases. \\ * - with limitations }
  \label{table:lvls_case_1-4}
  \centering
  \begin{tabular}{L{1.3in}C{0.8in}C{0.8in}C{0.8in}C{0.8in}}
  \hline
     Adoption level & Case I & Case II & Case III & Case IV \\

    \hline
      1~Org. readiness            & Current     & Current    & Current & Failed \\
      2~Component-level automation& Current     & Planned    & Planned* & Planned \\
      3~Solution-level automation & Planned*    & Planned*   & Planned* & Planned \\
      4~Continuous delivery       & Not feasible& Not feasible & Planned* & Planned \\
      5~Continuous data           & Potential*  & Potential* & Potential & Potential \\
      6~Evolving business         & Irrelevant  & Potential* & Potential & Potential \\
     
    \hline
 \end{tabular}
 \end{table*}

\subsubsection{Organizational support}

All studied companies, except Case IV, had strong management support and a clear vision for implementing the required changes. Importantly, Cases I - III perceived software as crucial for delivering their offerings. Thus, they were prepared to make the necessary investments and organizational changes to improve software delivery. 

In Case IV, an attempt to optimize and adopt more collaborative approach to software delivery stalled due to the lack of organizational support. Notably, this was the only case where software was not universally accepted as part of the core business and mostly outsourced to external vendors and consultants. Nevertheless, the organization wishes to leverage software to complement their core business with various software-intensive services. However, the conservative culture towards changes in the organization is yet to be altered.

The companies strive to reap the promised benefits of continuous delivery, yet the effort and adoption barriers often remain insurmountable. Notably, organizational inertia and conservative attitudes towards changes may hinder any innovation initiatives. Yet, we acknowledge that not every organization and type of business can benefit from CSE equally. As observed in Case IV, it remains to be seen if a chemical products company can sustain without treating software a key component of their business.

These findings are very much in line with literature on agile and CSE adoption. For instance, Neely et al.~\cite{neely2013continuous} highlights the organizational transformation as a crucial step in CSE adoption. Heeager et al.~\cite{heeager2018conceptual} lists executive support and compatible organizational support as key success factors in agile adoptions. 

Denning~\cite{denning2019ten} reports results from stalled agile transformations and reports that the lack of paradigm shift in organizations can hinder the success of the transformation. Specifically, unless the whole organization embraces the new ideas, any pockets of excellence are crushed by the surrounding bureaucracy.

In our earlier work, we presented a unified view of digital transformation and the adoption of continuous engineering
~\cite{klotins2022unified}. 
Both, CSE and digital transformation share similar conditions and outcomes and build upon each other. Namely, digital transformation sets the overall organizational strategy how to leverage software, while continuous engineering is a key mechanism to realize it. In tandem, a successful digital transformation and the adoption of continuous engineering lead to increased organizational performance.

In Case IV, attempts to adopt CSE may have been premature and preceded a broader organizational transformation towards digitalization. Kempeneer et al.~\cite{kempeneer2023virtual} reviews failed cases of digital transformation and categorizes barriers for successful transformations. Among them, a broad wave of changes in an organization creates a degree of chaos leading to further segregation, not unification, of different units. Without proper support and guidance this fortifies the silos stalling any unification efforts.

\paragraph{Insight I} The organizational readiness for CSE stems from a broader organizational strategy and commitment to digital transformation. Without a commitment and support to leverage software to transform the business, attempts to implement CSE may end in vain.

\subsubsection{Process automation initiatives}

All studied cases are looking to improve internal efficiency by adopting automation and elements of continuous integration and internal delivery. Automating build, integration, and test processes is relatively straightforward. Yet, our results highlight several caveats. See Table~\ref{table:lvls_case_1-4}, Levels 2-3.

Firstly, for large and safety critical products, verification is a complicated process and only parts of this process can be automated. As in Cases I and II, the software vendor controls only part of the delivery pipeline and the final responsibility of installing and verifying the product lies with someone else. Importantly, the final verification of an industrial automation solution or an automotive component considers the system as a whole. That is, the final verification asserts safe operation of a facility or a vehicle as a whole according to regulations, not that the individual components requirements are met. Hence, the software vendor cannot fully automate this process in-house. In Cases I and II, companies addressed this challenge with reference solutions or test setups mimicking real life installations of their products. Moving working software from test rigs to actual deployment for use still remains a largely plan-driven and manual process.

Secondly, a dense network of various cross-organizational dependencies limit how much a single team can accomplish without involving other organizational units. As in Case III, in a large organization the adoption of new ways of working is not homogeneous. Pockets of pioneers and laggards must co-exist and still deliver working software. Such a situation is fragile as either the organizations embraces change and the new ways of working prevail or the initiatives die off due to lack of broader support~\cite{denning2019ten}.

Thirdly, as demonstrated in Case IV, adopting new ways of working may require new roles and render some roles obsolete.
If the organization is not prepared to follow through and accomodate the changes, the transformation process may end up stalling. Moreover, the adoption of CSE may require the organization to take control of previously outsourced functions, creating further disruptions and requiring a radical transformation~\cite{cichosz2020digital}.

\paragraph{Insight II} During CSE adoption, pockets of laggards and pioneers must coexist, follow a coherent and compatible process, and produce working software. Special care must be paid to ensure the good initiatives survive and laggards are encouraged to adopt the new ways of working.

\subsubsection{Towards data-driven engineering}

Interestingly, none of the experts explicitly mentioned data and data-driven engineering as aims for adopting CSE in any of the cases. In cases I-II, companies does not have access to their products after decommissioning. Hence, accessing product usage data and using it to guide further product development is out of question. Yet, also Cases III-IV where access to data and telemetry is possible, makes no mention of benefiting from data.

State-of-the-art emphasizes the potential from aggregating and analyzing volumes of data to steer innovation, drive change, and increase value delivery~\cite{gunther2017debating}. Access and analysis of data is listed as one of the preconditions for digital transformation~\cite{vial2021understanding}. The ability to collect a more fine-grained data is mentioned as a key benefit from adopting CSE~\cite{humble2018accelerate}. 

This finding also illustrates an overlooked opportunity to collect at least some data to drive product and process decisions. Understandably, not every organization can rely on a constant stream of product usage data. Yet, data quantifying internal process performance and asset quality  should be easier to access; see, for example, SPACE~\cite{forsgren2021space}, DORA metrics~\cite{wilkes2023framework}, Zabardast et al.~\cite{zabardastexploring} and Paudel et al.~\cite{paudel2024data}.

We argue that the real efficiency gains in an organization stem from data-driven continuous improvements rather than a big-bang blanket policy of automating everything~\cite{birk2001practical}. Therefore, the attempts to implement CSE should start by deploying instrumentation to collect at least some data to inform further improvements.

\paragraph{Insight III} Continuous improvement fueled by both internal and external data, and fast feedback loops is an overlooked benefit of CSE.

\subsubsection{Evolving business models}

All studied cases are relatively mature and established companies with stable business models. Yet, most identify the opportunity to evolve their businesses due to external pressures. However, changing a business model must happen in harmony with the rest of their respective ecosystems. 

In Case II, the automotive domain is shifting towards software-controlled driving experiences and OEM automotive brands wish to control the experience. Hence, the control of driving experience is lifted from individual components to a central computer. On the one hand, this simplifies the components, on the other hand, tweaking the driving experience requires a closer collaboration between suppliers and OEM companies. This opens up opportunities for more collaborative business models between OEM's and the suppliers.

In Case III, the global retailer is forgoing a shift from physical to digital shopping experiences. Rapid iterations, continuous improvement, and fine-tuning the customer experiences can be enabled with CSE. However, Case III is also intertwined in a complex partner and supplier network each with their own IT-systems who not necessarily share the same ambitions. 

The company behind Case IV also realizes the potential to add software-enabled services on top of their core chemistry business. However, the company is in the early stages of promoting software to the core of their value delivery chain. 

The evolution of business models in all the cases is slow. The primary reason being that one team, department, or even a company cannot force other neighboring units to transform at the same pace. Besides, these other units may face their own special constraints limiting the pace of the overall ecosystem evolution.

\paragraph{Insight IV} Even relatively stable and mature domains and business models evolve, albeit slowly.

\subsection{RQ2: What driving forces and constraints determine the optimal CICD adoption level?}


The primary reasons for considering CSE are related to software delivery efficiency and quality. The adoption constraints stem from the organizational context, type of product, and the supply chain. 

\subsubsection{Internal efficiency}

The companies in our study aim to automate software delivery's build, test, and integration steps to offload repetitive tasks from developers (Cases I-III). 
In Case IV, the aim is to consolidate and optimize IT operations. Currently, most software development is distributed to consultants. Software operations are handled in-house. Such a divide creates unnecessary complexities in software deployment and maintenance. Taking over at least some of the development and introducing DevOps practices is a way to reduce the complexity and streamline the operations.


\subsubsection{Improving quality and preventing defect slippage}
Another reason for introducing CSE is to improve quality by detecting defects early. In a plan-driven model, defects are often discovered late in the process when it is expensive to fix them, and any delays may cause slipping deadlines downstream~\cite{damm2006using}. With the introduction of automation, companies hope to shorten the release cycle and improve the feedback loop between developers and quality assurance.















\subsubsection{CSE adoption constraints}

Our results point to several interlinked constraints limiting the adoption of CSE by the book; see, for example, Forsgren et al.~\cite{humble2018accelerate}. Importantly, in all the cases, we identified several adoption constraints, hence it is difficult to pinpoint one key reason for the difficulties of CSE adoption.  We organize these constraints into three groups: organization, product, and supply chain dependencies.

\textit{Organization} related constraints comprise inertia (Cases II-IV), lack of management support, and conservative attitudes (Cases III-IV). In all studied cases, organizational change takes time, hence the adoption of CSE is gradual. 

As illustrated by Case IV, organizational change may involve taking over previously outsourced functions, introducing new ones, and obsoleting irrelevant roles and skills. Furthermore, in Case IV, there is a lack of management support for the required change, thus the CSE adoption had collapsed altogether. 
In Case III, the CSE adoption is slowed down by a dense and complex network of dependencies. One single team or unit may not have enough leverage to push others to adapt. 

\textit{Product} related constraints comprise product composition, and regulatory requirements. For instance, in Cases I and II, the products contain both hardware and software components; thus, software iterations should be synced with hardware changes. 

In Case I, the product and its deployment is highly regulated, thus limiting how much can be controlled by the vendor and automated. Once installed, the whole facility must forgo verification - a complex procedure outside software vendors control.

\textit{Supply chain} related constraints pertain to upstream and downstream dependencies on the product and the organization. 
In Cases I - II, the product is passed to another organization for installation and maintenance. Therefore, the supplier has limited control over the process and must adhere to the expectations of the receiving side. 
Similarly, in Case III, the software is created collaboratively by many teams who depend on each other for inputs and outputs. Any change needs to be agreed between the involved units.

\textit{Other} While our experts recognize the technology constraints, they do not seem to be the primary adoption barriers. Rather, the technology constraints are challenging to remove due to organizational barriers.

\paragraph{Insight V} Most CSE adoption constraints are related to organizational inertia and positioning in the market, rather than technology.













\subsection{RQ3: To what extent is the CSE industry readiness model useful for setting goals and identifying constraints in the adoption of CSE?}

The aim of the CSE Industry Readiness Model
~\cite{klotins2023continuous} 
is to aid practitioners in setting realistic expectations and foresee pitfalls in CSE adoption efforts. In this study, we use the CSE Industry Readiness Model to analyze the cases and compare model predictions with experts' assessments. In this section, we discuss the lessons learned and suggest improvements to the readiness model.

A typical way to evaluate a model is to compare the model's predictions with empirical observations~\cite{foss2003simulation,pineiro2008evaluate}. We summarize predicted and observed adoption levels in Tables~\ref{table:lvls_case_1}-\ref{table:lvls_case_4}. 

\subsubsection{Level 1 - Organizational readiness}
The model suggests the foundation and first adoption level is to accomplish organizational readiness for CSE. The readiness encompasses a relative stability of the product and the business model, sufficient resources and management support, as well as minimization of organizational silos~\cite{klotins2023continuous}.

Companies have successfully attained readiness in Cases I-III, both by our evaluation and observation. In Case IV, the CSE adoption effort was abandoned due to the lack of management commitment to follow through with necessary changes. Moreover, in Case IV, the software engineering effort was divided between various external parties with limited communication, hence creating silos. Hence, we conclude that the readiness model correctly identifies organizational readiness as the first building block toward CSE adoption.

We identify that the readiness model does not highlight the market and regulatory factors to a sufficient degree. The distinction between internal and external barriers is important to set the scope for implementing CSE.

\subsubsection{Level 2 - Component-level automation}
Level 2 of the industry readiness model suggests that the adoption of CSE starts with limited, component-level automation initiatives. The introduction of automation requires a basic source code version control system infrastructure, building and testing infrastructure, minimizing bottlenecks such as manual steps, and delivering large, sweeping changes at once~\cite{klotins2023continuous}.

In our study, except in Case IV, we observe that this step is largely achieved, both by expert observation and prediction of the model. Regardless of CSE adoption, automation has become prevalent in software development~\cite{soares2022effects}. In practice, there are a few limitations to implementing contained automation initiatives.

In Case IV, automation initiatives are difficult to enforce as most of the engineering effort is handled by external vendors outside the company's direct control. Unless the organizational readiness is achieved and the company takes control of the development, or require 3rd party vendors to follow continuous engineering practices, achieving level 2 is not feasible. 

The observation of the experts matches The Industry Readiness Model prediction - the adoption of automation starts with isolated initiatives.

\subsubsection{Level 3 - Solution-level automation}
Level 3 of the readiness model, Solution-level automation attempts to connect different isolated automation 
initiatives into a coherent pipeline allowing end-to-end building and verification of the solution~\cite{klotins2023continuous}.. 

Our study shows that solution level automation extends component-level automation, just as suggested by the model. 
Furthermore, the model hints that a degree of alignment and agreement between parties is required to establish an automated pipeline. As illustrated by Cases I-II, synchronizing hardware and software development cycles, further synchronizing development cadence up and downstream is a major undertaking outside the purview of any individual unit or company. 

The experiences from our study participants matches the model predictions about Level 3 readiness.

\subsubsection{Level 4 - Continuous delivery}

Although none of the studied cases have achieved Level 4, the lessons learned in this study supports the prediction that access to production environments and a service level agreement permitting automated software deliveries are the major concerns in establishing continuous delivery practices~\cite{klotins2023continuous}..

We argue that unrestricted access to deploy software automatically is an exception on business-to-business domain. However, access to internal process data is much more feasible. Therefore, we identify that the model needs to separate internal and external data collection.

\subsubsection{Level 5 - Continuous feedback}
Level 5 in the CSE readiness model predicts that combining data from development automation pipelines (Levels 2-3) and continuous delivery (Level 4) enables a feedback loop to support product planning and fine-tuning~\cite{klotins2023continuous}.

In this level we observe discrepancies between expert observations and model predictions. Continuous improvement addresses two separate improvement areas - improvements in \textit{HOW} the product is developed, and improvements in \textit{WHAT} is being delivered.  

Most organizations, with an exception of Case IV, control their development process, can collect metrics about their software delivery performance, and make the necessary improvements, given enough time and support. Hence, it is useful to consider the internal improvements on how the software is developed separately.

The studied cases illustrate the practical constraints in collecting product usage data and using such data to fine-tune the product. Moreover, as in Case I, customers are interested in the stability and robustness of the current product version, not the new features. That being said, the lack of access to product usage data should not stop the organizations to improve their internal efficiency. Hence, we deem necessary to separate the two improvement loops in the model.

\subsubsection{Level 6 - Evolving business models}

The industry readiness model~\cite{klotins2023continuous} implies that continuous feedback along with end-to-end automation enables organizations to evolve their business models. However, our results show that the business model evolution is driven more with general trends in the industry than with the internal capability for evolution. 

In Case II, the automotive business models are evolving due to the need to provide a more consistent driving experience. In Case 3, the retail business model is forgoing a shift towards online experiences. Case 4 could potentially consider adding software-intensive services on top of their core chemistry business. Therefore, we conclude that shifts in business models are driven by external factors and made possible by internal improvements in digitalization, automation, and continuous improvement.

\subsubsection{The Updated Model}

\begin{figure*}[ht!]
  \centering
  \includegraphics[width=\textwidth]{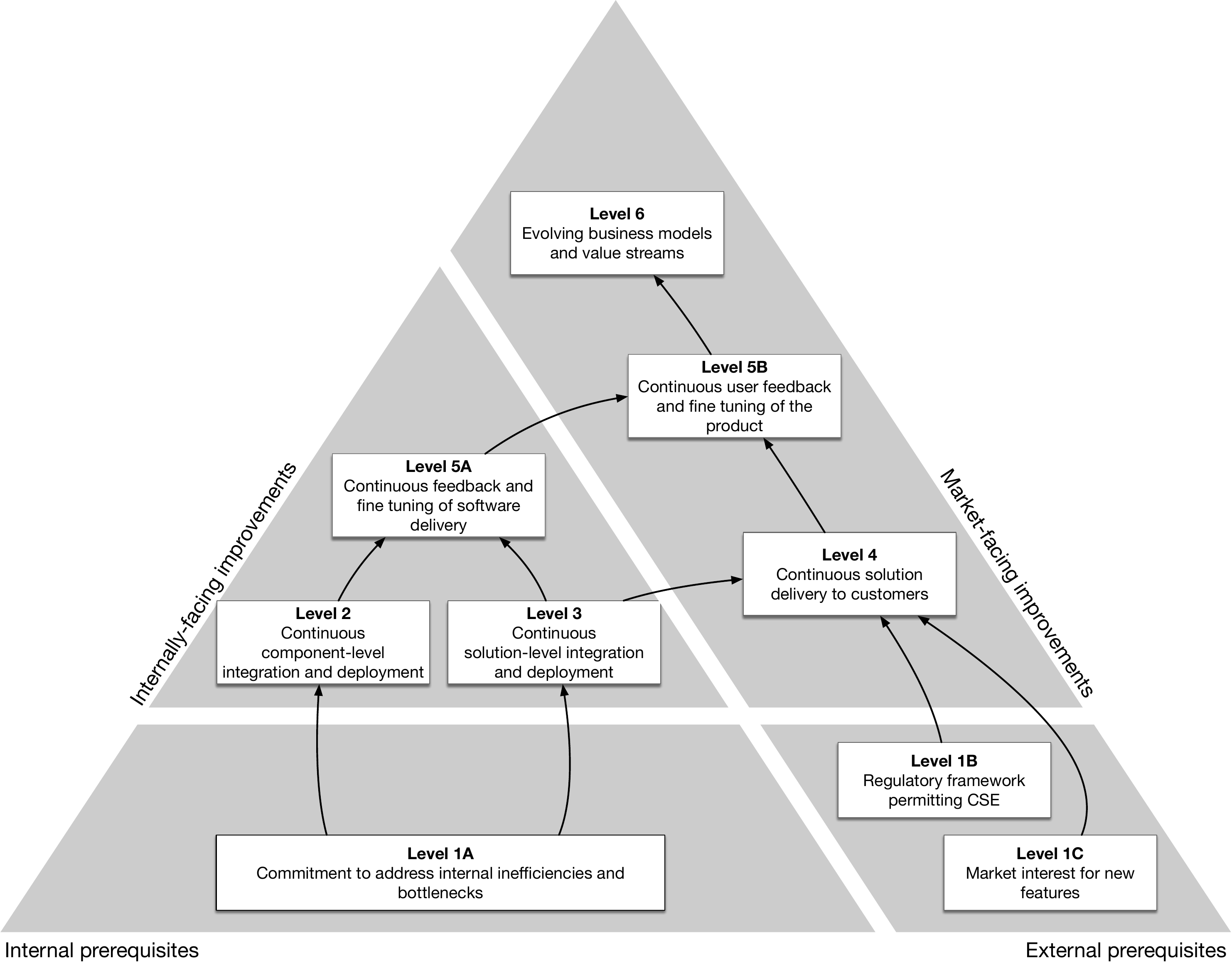}

  \caption{The updated industry readiness model.}
  \label{fig:rediness_model_updated}     
\end{figure*}

The industry readiness model~\cite{klotins2023continuous} largely matches with the expert observations. However, as discussed in earlier sections, we see an opportunity for imprving the model. Specifically, we propose the following improvements and present the updated model in Fig.~\ref{fig:rediness_model_updated}.

\textit{Separate Level 1 into readiness for internal improvements, market interest for frequent new features, and suitable regulatory framework for CSE}. As suggested by the cases, many obstacles to end-to-end adoption of CSE stem from domain conventions or the market situation. Hence, we separate the two underlining that internal improvements are still possible despite market or domain constraints.

\textit{Separate Level 5 into internal and external improvement loops}. Splitting Level 5 into levels 5A and 5B would cover potential scenarios where a company has access to internal software delivery metrics and can follow continuous improvement to fine-tune \textit{HOW} software is developed (Level 5A). Hence, the Level 5B cover using telemetry from the software in use to support product planning.

\textit{Introduce areas of concern}. Our analysis show that the path to CSE adoption is hindered by four areas of concern. Namely, meeting internal readiness, market and domain suitability for CSE, internal improvements in adopting automation, and improvements concerning continuous software delivery to market. We deem useful to include these areas in the model.

Internal prerequisites concern internal readiness and commitment to follow through with the CSE adoption initiatives (as in Cases I--III, and Case IV as counterexample). External prerequisites denote market factors, such as whether customers deem immediate access to new features valuable (as in Cases I--III, and supply chain concerns (as in Case II). Internally-facing improvements comprise streamlining software delivery steps within the organization. Market-facing improvements group together CSE steps concerning the interface with customers and markets.

Using areas of concern allows quickly pinpoint the likely CSE adoption target. For instance, in Cases I--II, the internal prerequisites are achieved, internally-facing improvements are under way, however market facing improvements are not relevant at the time due to external factors. In Case III, both internal and external prerequisites are met, and it is feasible to consider both internal and external improvements to adopt CSE. In Case VI, due to missing internal prerequisites, the CSE adoption has stalled altogether.

In the figure, see Fig.~\ref{fig:rediness_model_updated}, white boxes show CSE adoption levels. The arrows between the boxes denote likely paths and dependencies between the levels, e.g. to achieving continuous delivery to customers (Level 4), requires both market and domain readiness (Level 1B) and internal capability for continuous solution level deployment (Level 3). The gray background areas denote the areas of concern.

\section{Conclusions}

In this paper we present and discuss four cases of introducing CSE in regulated and otherwise complex domains. Our findings are based on four expert testimonies. The experts who have significantly contributed in writing this paper were involved in the studied cases, therefore providing rich insights and interpretations on their CSE adoption efforts. We present the cases as experience reports and analyze the state-of-practice in CSE adoption, adoption driving forces, challenges, and attempt to devise an improved CSE readiness and adoption model. 

Our findings are aligned with earlier studies suggesting that large organizational transformations are challenging. Eliminating old and introducing new ways of working are met with resistance from employees and customers, legacy software and organizational structures are difficult to change overnight, thus pockets of innovators and laggards must co-exist all while delivering working software and ensuring business continuity. 

The most significant bottleneck for CSE adoption is market and domain readiness. Established companies operate within complex ecosystems of suppliers, customers, partners, and regulators. While these ecosystems evolve and new trends emerge, no single company can dictate terms for the others.

As a result, full (Level 6) CSE adoption in complex domains will likely be gradual, constrained by practical considerations such as domain characteristics, product types, regulatory requirements, and customer preferences.

In the meantime, organizations can focus on partial adoption of CSE practices and focus in internally-facing improvements. Those are not restricted by external factors and can yield significant improvements in software delivery efficiency and preparedness for future trends in software delivery in their respective domains.

\section*{Acknowledgements}
This research is supported by Swedish Knowledge Foundation (KK-stiftelsen), and project SERT Profile (Ref. 2018/010).


\bibliography{bibliography}

\end{document}